\documentclass[preprintnumbers, prl, twocolumn, showpacs, floatfix,
preprintnumbers,  superscriptaddress, nofootinbib]{revtex4-1}
\usepackage[dvips]{graphicx}
\usepackage{epsf}
\usepackage{amsmath}
\usepackage{amssymb}
\usepackage{graphicx}
\usepackage{dcolumn}
\usepackage{bm}
\usepackage{color}
\usepackage{epsfig}
\usepackage{amsfonts}
\usepackage{ulem} 

\def\be{\begin{equation}}
\def\ee{\end{equation}}
\def\ba{\begin{eqnarray}}
\def\ea{\end{eqnarray}}

\def\be{\begin{equation}}
\def\ee{\end{equation}}
\def\bea{\begin{eqnarray}}
\def\eea{\end{eqnarray}}

\newcommand{\ud}{\mathrm{d}}

\begin{document}

 \title{``Teleparallel'' Dark Energy}

\author{Chao-Qiang Geng}
\email{geng@phys.nthu.edu.tw}
\affiliation{Department of Physics, National Tsing Hua University,
Hsinchu, Taiwan 300}
 \affiliation{National Center for Theoretical Sciences, Hsinchu,
Taiwan 300}

\author{Chung-Chi Lee}
\email{g9522545@oz.nthu.edu.tw}
 \affiliation{Department of Physics, National Tsing Hua University,
Hsinchu, Taiwan 300}

\author{Emmanuel N. Saridakis}
\email{Emmanuel_Saridakis@baylor.edu}
 \affiliation{CASPER,
Physics Department,
Baylor University,
Waco, TX  76798-7310, USA}
\affiliation{National Center for Theoretical Sciences, Hsinchu,
Taiwan 300}

\author{Yi-Peng Wu}\email{s9822508@m98.nthu.edu.tw}
 \affiliation{Department of Physics, National Tsing Hua University,
Hsinchu, Taiwan 300}


\begin{abstract}
Using the ``teleparallel'' equivalent of General Relativity as the
gravitational sector, which is based on torsion instead of
curvature, we add a canonical scalar field, allowing for a
nonminimal coupling with gravity. Although the minimal case is
completely equivalent to standard quintessence, the nonminimal
scenario has a richer structure, exhibiting quintessence-like or
phantom-like behavior, or experiencing the phantom-divide
crossing. The richer structure is manifested in the absence of a
conformal transformation to an equivalent minimally-coupled model.
\end{abstract}

\maketitle

\section{Introduction}

The ``teleparallel'' equivalent of General
Relativity (TEGR) \cite{ein28,Hayashi79} is an equivalent
formulation of classical gravity, in which instead of using
the torsionless Levi-Civita connection one uses the
curvatureless Weitzenb{\"o}ck one. The dynamical objects are the
four linearly
independent vierbeins (these are \emph{parallel} vector fields
represented by the appellation ``teleparallel''). The
advantage of this framework is
that the torsion tensor is formed solely from products of first
derivatives of the tetrad. As described in \cite{Hayashi79}, the
Lagrangian density $T$ can
be constructed from this
torsion tensor under the assumptions of invariance under general
coordinate transformations, global Lorentz transformations, and
the parity operation, along with requiring the Lagrangian density
to be second order in the torsion tensor. Thus, apart from possible
conceptual differences, TEGR is completely equivalent and indistiguishable
form General Relativity (GR) at the level of equations, both background
and perturbation ones.

On the other hand, in General Relativity one can
add the quintessence scalar field in order to acquire a
dynamical dark energy sector, a scenario that exhibits a very
interesting cosmological behavior and has gained a huge amount of
research \cite{quintessence}. Amongst others, one can generalize it
by including a nonminimal coupling between the quintessence field and gravity
\cite{nonminimal}, or more generally extend it to the scalar-tensor
paradigm \cite{Maedabook}. One can also use a phantom instead of a
canonical field \cite{phant}, or the combination of both these fields in a
unified scenario called quintom \cite{quintom}.

In this letter we are interested in formulating ``teleparallel
dark energy'', adding a canonical scalar field in TEGR. In the
minimal coupling case the resulting theory is identical to the ordinary
quintessence, both at the background and perturbation levels.
However, when the nonminimal coupling is switched
on, teleparallel dark energy is different from its
GR counterpart, and the cosmological behavior of such a new
scenario proves to be very interesting.

\section{Quintessence in General Relativity}\label{quintGR}

Let us
review very briefly the quintessence paradigm in General
Relativity. In such a scenario the dark energy
sector is attributed to a homogeneous scalar field $\phi$, and the action
is given by~\cite{nonminimal}
\begin{equation}
S=\int\ud^{4}x\sqrt{-g}\Bigg[\frac{R}{2\kappa^{2}}
+ \frac{1}{2} \Big(\partial_{\mu}\phi\partial^{\mu}\phi+\xi
R\phi^{2}\Big) - V(\phi)+\mathcal{L}_m\Bigg] \label{action},
\end{equation}
with  $\kappa^{2}=8\pi G$,
$c=1$,
 $V(\phi)$ the scalar-field potential,
$\xi$ the non-minimal coupling parameter, $R$ the Ricci scalar, and
$\mathcal{L}_m$ the matter Lagrangian. Note
the difference in the metric signature that exists amongst the various
works in the literature and the corresponding sign changes in the action,
since a change in the metric signature leads $g_{\mu\nu}$,
$\Box$ and $R_{\mu\nu}$ to change sign, while $R$ and the energy-momentum
tensor remain unaffected \cite{nonminimal}. In this letter
we use the signature $(+,-,-,-)$ in all sections,
just to be closer to the
literature of teleparallel gravity. Thus, under this convention, the
conformal value of $\xi$ is
$-1/6$.

In the case of
a flat Friedmann-Robertson-Walker (FRW) background metric
\begin{equation}
\label{FRWmetric}
ds^2= dt^2-a^2(t)\,\delta_{ij} dx^i dx^j,
\end{equation}
where $t$ is the cosmic time, $x^i$ are the comoving spatial
coordinates and $a(t)$ is the scale factor,
the Friedmann equations write as
\begin{eqnarray}
\label{FR1}
H^{2}=\frac{\kappa^2}{3}\Big(\rho_{\phi}+\rho_{m}\Big),
\dot{H}=-\frac{\kappa^2}{2}\Big(\rho_{\phi}+p_{\phi}+\rho_{m}+p_{m}
\Big),~~~~
\end{eqnarray}
where $H=\dot{a}/a$ is the Hubble parameter,
a dot denotes
differentiation with respect to $t$, and
 $\rho_m$ and $p_m$ are the matter energy density and
 pressure, respectively,  following the standard evolution equation
$\dot{\rho}_m+3H(1+w_m)\rho_m=0$, with $w_m= p_m/\rho_m$
the matter equation-of-state parameter. Additionally, we have introduced
the energy density and pressure of the nonminimally coupled scalar field,
given by \cite{nonminimal}:
\begin{eqnarray}
\label{rhophi}
 &&\rho_{\phi}=  \frac{1}{2}\dot{\phi}^{2} + V(\phi)
-
  6 \xi H\phi\dot{\phi} -  3\xi H^{2}\phi^{2},\\
 &&p_{\phi}=  \frac{1}{2}(1+4\xi)\dot{\phi}^{2} - V(\phi)  + 2 \xi
(1+6\xi)\dot{H}\phi^{2}  \nonumber\\
 \label{pphi}
 &&  \ \ \ \ \  -   2 \xi
H \phi\dot{\phi}
+  3\xi(1+8\xi)H^{2}\phi^{2} - 2\xi\phi V'(\phi),
\end{eqnarray}
where a prime denotes derivative with respect to $\phi$. We note that
the
above relations have been simplified by using the useful expression
$R=6(\dot{H}+2H^2)$
in the FRW geometry.

As  mentioned above, in such a scenario, dark energy is attributed to
the scalar field, and thus its equation-of-state  parameter reads:
\begin{equation}\label{EoS}
w_{DE}\equiv w_\phi=\frac{p_\phi}{\rho_\phi}.
\end{equation}
 Finally, the equations close by considering the
evolution equation for the scalar field
\cite{nonminimal}:
\begin{equation}
\ddot{\phi}+3H\dot{\phi}-6\xi(\dot{H}+2H^2)\phi+   V'(\phi)=0,
\end{equation}
which can alternatively be written in the standard form
$
\dot{\rho}_\phi+3H(1+w_\phi)\rho_\phi=0$.

\section{Teleparallel Equivalent to General Relativity (TEGR) }
\label{TEGR}

We now briefly review TEGR. The
notation is as
follows: Greek indices $\mu, \nu,$...
and capital Latin indices $A, B, $...
run over all coordinate and tangent
space-time 0, 1, 2, 3, while lower case Latin indices (from the middle of the
alphabet) $i, j,...$ and lower case Latin indices
(from the beginning of the alphabet) $a,b, $...
run over spatial and tangent
space coordinates 1, 2, 3, respectively.

As  stated in {\em Introduction}, the dynamical variable of
``teleparallel'' gravity is the vierbein
field ${\mathbf{e}_A(x^\mu)}$. This forms an orthonormal basis
for the tangent
space at each point $x^\mu$ of the manifold, that is $\mathbf{e}
_A\cdot%
\mathbf{e}_B=\eta_{AB}$, where $\eta_{AB}=diag (1,-1,-1,-1)$.
Furthermore,
the vector $\mathbf{e}_A$ can be analyzed with the use of its
components $%
e_A^\mu$ in a coordinate basis, that is
$\mathbf{e}_A=e^\mu_A\partial_\mu $.

In such a construction, the metric tensor is obtained from the
dual vierbein
as
\begin{equation}  \label{metrdef}
g_{\mu\nu}(x)=\eta_{AB}\, e^A_\mu (x)\, e^B_\nu (x).
\end{equation}
Contrary to GR, which uses the torsionless
Levi-Civita
connection, in TEGR ones takes the curvatureless
Weitzenb%
\"{o}ck connection \cite{Weitzenb23}, whose torsion tensor reads
\begin{equation}  \label{torsion2}
{T}^\lambda_{\:\mu\nu}\equiv \overset{\mathbf{w}}{\Gamma}^\lambda_{
\nu\mu}-%
\overset{\mathbf{w}}{\Gamma}^\lambda_{\mu\nu}
=e^\lambda_A\:(\partial_\mu
e^A_\nu-\partial_\nu e^A_\mu)\,,
\end{equation}
where $\overset{\mathbf{w}}{\Gamma}^\lambda_{\nu\mu}\equiv e^\lambda_A\: \partial_\mu
e^A_\nu$.
Moreover, the contorsion tensor, which equals to the difference
between Weitzenb%
\"{o}ck and Levi-Civita connections, is defined as
$K^{\mu\nu}_{\:\:\:\:\rho}\equiv-\frac{1}{2}\Big(T^{\mu\nu}_{
\:\:\:\:\rho}
-T^{\nu\mu}_{\:\:\:\:\rho}-T_{\rho}^{\:\:\:\:\mu\nu}\Big)$  and we also
define
$
S_\rho^{\:\:\:\mu\nu}\equiv\frac{1}{2}\Big(K^{\mu\nu}_{\:\:\:\:\rho}
+\delta^\mu_\rho
\:T^{\alpha\nu}_{\:\:\:\:\alpha}-\delta^\nu_\rho\:
T^{\alpha\mu}_{\:\:\:\:\alpha}\Big)$.

In the present formalism all the information
concerning the
gravitational field is included in the torsion tensor
${T}^\lambda_{\:\mu\nu} $. Using the above quantities one can extract the
form of the ``teleparallel Lagrangian'', which is nothing else than the
torsion scalar, namely \cite{ein28,Hayashi79,Maluf:1994ji}:
{\small{
\begin{equation}  \label{telelag}
\mathcal{L}=T\equiv
S_\rho^{\:\:\:\mu\nu}\:T^\rho_{\:\:\:\mu\nu}=\frac{1}{4}T^{\rho
\mu \nu }T_{\rho \mu \nu }+\frac{1}{2}T^{\rho \mu \nu
}T_{\nu \mu \rho }-T_{\rho \mu }^{\ \ \rho }T_{\ \ \ \nu }^{\nu
\mu }.
\end{equation}}}

In summary, the simplest action in a universe governed by teleparalel
gravity is
\begin{eqnarray}  \label{action}
I =\int d^4x e
\left[\frac{T}{2\kappa^2}+\mathcal{L}_m\right],
\end{eqnarray}
where $e = \text{det}(e_{\mu}^A) = \sqrt{-g}$ (one could also include a
cosmological constant).
Variation with respect to the vierbein fields gives equation of motion
\begin{equation}\label{eom}
e^{-1}\partial_{\mu}(ee_A^{\rho}S_{\rho}{}^{\mu\nu})
-e_{A}^{\lambda}T^{\rho}{}_{\mu\lambda}S_{\rho}{}^{\nu\mu}
-\frac{1}{4}e_{A}^{\nu
}T
= \frac{\kappa^2}{2}e_{A}^{\rho}\overset {\mathbf{em}}T_{\rho}{}^{\nu}, \
\end{equation}
where
$\overset{\mathbf{em}}{T%
}_{\rho}{}^{\nu}$ stands for the usual
energy-momentum tensor.
These equations are exactly the same as those of GR for every geometry
choice.
In particular, for the FRW background metric (\ref{FRWmetric}),
the vierbein choice of the form 
\begin{equation}  \label{FRWvierbeins}
e_{\mu}^A=\mathrm{diag}(1,a,a,a)
\end{equation}
is an exact solution~\cite{Hayashi79} of the field equation in Eq.~(\ref{eom}),
which does not generate a divergent energy for the whole space-time.
Furthermore,  it is easily seen that the corresponding Friedmann equations are
identical to the GR ones, both at the background and
perturbation levels \cite{ein28,Hayashi79,Maluf:1994ji}.

\section{Teleparallel Dark Energy}
\label{TEGRquint}

Let us now construct teleparalell dark energy. This will be done by
adding a scalar field in the equivalent, teleparallel, formulation of GR.
Thus, the
action will simply read:
\begin{equation}
S=\int\ud^{4}x e\Bigg[\frac{T}{2\kappa^{2}}
+ \frac{1}{2} \Big(\partial_{\mu}\phi\partial^{\mu}\phi+\xi
T\phi^{2}\Big) - V(\phi)+\mathcal{L}_m\Bigg]. \label{action2}
\end{equation}
We emphasize that in the above action  a nonminimal coupling between the
scalar field and gravity is allowed. Although in the nonminimal case one
could use alternative torsion scalars, we prefer to keep the standard one
for simplicity.  We also note that the action in (\ref{action2}) with the
torsion formulation of GR is similar to the standard nonminimal
quintessence where the scalar field couples to the Ricci scalar.

Variation of action (\ref{action2}) with respect to the vierbein fields
yields equation of motion
\begin{eqnarray}\label{eom2}
\left(\frac{2}{\kappa^2}+2 \xi
\phi^2 \right)\left[e^{-1}\partial_{\mu}(ee_A^{\rho}S_{\rho}{}^{\mu\nu} )
-e_{A}^{\lambda}T^{\rho}{}_{\mu\lambda}S_{\rho}{}^{\nu\mu}
-\frac{1}{4}e_{A}^{\nu
}T\right]\nonumber\\
-
e_{A}^{\nu}\left[\frac{1}{2}
\partial_\mu\phi\partial^\mu\phi-V(\phi)\right]+
  e_A^\mu \partial^\nu\phi\partial_\mu\phi\ \ \nonumber\\
+ 4\xi e_A^{\rho}S_{\rho}{}^{\mu\nu}\phi
\left(\partial_\mu\phi\right)
=e_{A}^{\rho}\overset {\mathbf{em}}T_{\rho}{}^{\nu}.~~~~~~~~~~~~
\label{eom2}
\end{eqnarray}
Therefore, imposing the FRW geometry of the form (\ref{FRWvierbeins}) (that
is
(\ref{FRWmetric})) we obtain the same Friedmann equations as in
the conventional quintessence, namely
(\ref{FR1}),
however in this case the scalar field energy density and pressure become:
\begin{eqnarray}
\label{telerho}
 &&\rho_{\phi}=  \frac{1}{2}\dot{\phi}^{2} + V(\phi)
-  3\xi H^{2}\phi^{2},\\
 &&p_{\phi}=  \frac{1}{2}\dot{\phi}^{2} - V(\phi) +   4 \xi
H \phi\dot{\phi}
 + \xi\left(3H^2+2\dot{H}\right)\phi^2.\ \ \ \ \
 \label{telep}
\end{eqnarray}
Additionally, variation of the action with respect to the scalar field
provides its evolution equation, namely:
\begin{equation}
\ddot{\phi}+3H\dot{\phi}+6\xi H^2\phi+   V'(\phi)=0.
\label{fieldevol2}
\end{equation}
Note that in the above expressions we have used the useful relation
$T=-6H^2$, which straightforwardly arises from the calculation of
(\ref{telelag})
for the FRW geometry.

In this scenario, similar to the standard quintessence, dark
energy is attributed to the scalar field, and thus its equation-of-state
parameter ($w_{DE}$) is defined to be the same as that in (\ref{EoS}),
but $\rho_\phi$ and $p_\phi$ are now given by (\ref{telerho}) and
(\ref{telep}), respectively.
 Finally, one can see that the scalar field evolution (\ref{fieldevol2})
 leads to the standard relation
$\dot{\rho}_\phi+3H(1+w_\phi)\rho_\phi=0$.

\section{Cosmological Implications}\label{implications}

We
now explore the cosmological implications of the scenario at hand.
Firstly, we immediately observe that in the case of the minimal coupling,
teleparallel dark energy coincides with quintessence (see
(\ref{rhophi})-(\ref{pphi}) and (\ref{telerho})-(\ref{telep})), and one
can verify that at the level of perturbations too. This is expected since,
concerning the gravitational sector, TEGR is identical with GR, and in the
minimal case one just adds a distinct scalar
sector, thus making no difference whether it is added in either of the two
theories. However, things are different if we switch on the nonminimal
coupling. In this case the additional scalar sector is coupled to gravity,
with the curvature scalar in GR and with the torsion scalar in TEGR, and
thus the resulting coupled equations do not coincide. Clearly,
teleparallel dark energy, under the nonminimal coupling, is a different theory.

Let us proceed in presenting some basic and general features of
the
nonminimal coupling of the scalar-torsion theory.
Apart from the straightforward results that
 dark energy possesses a dynamical nature as well as it can
drive the universe acceleration, the most interesting and direct
consequence of the dark energy density and pressure relations
(\ref{telerho})-(\ref{telep}) is that the dark energy
equation-of-state parameter can lie in the quintessence regime
($w_{DE}>-1$), in the phantom regime ($w_{DE}<-1$), or exhibit the
phantom-divide crossing during cosmological evolution. This is a
radical difference with the quintessence scenario and reveals the
capabilities of the construction.

In order to present the above features in a more transparent way,
we evolve numerically the cosmological system for dust matter
($w_m\approx0$), using the redshift $z=a_0/a-1$ as the independent
variable, imposing the present scale factor $a_0$ to be equal to
1, the  dark energy density
$\Omega_{DE}\equiv\kappa^2\rho_{\phi}/(3H^2)$ at present to be
$\approx0.72$ and its initial value to be $\approx0$. Finally,
concerning the scalar field potential we use the exponential
ansatz of the form $V=V_0e^{\lambda\phi}$.
\begin{figure}[!]
\mbox{\epsfig{figure=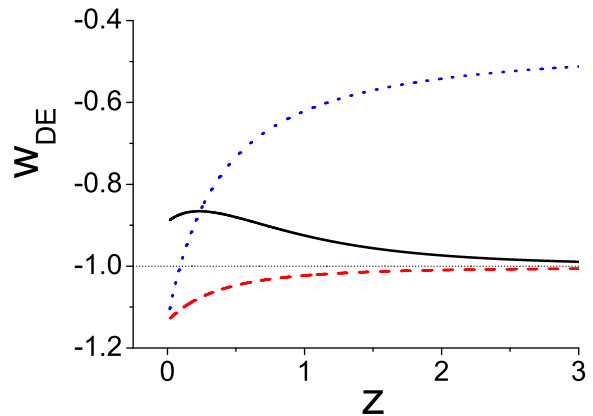,width=8.74cm,angle=0}} \caption{{\it
Evolution of the dark energy equation-of-state parameter $w_{DE}$
as a function of the redshift $z$, for three cases of the
teleparallel dark energy scenario, in the exponential scalar-field
potential ansatz of the form $V=V_0e^{\lambda\phi}$. The
black-solid curve presents quintessence-like behavior and
corresponds to $\xi=-0.4$, $\lambda=1.5$ and $V_0\approx2\times
10^{-13}$, the red-dashed curve presents phantom-like behavior and
corresponds to $\xi=-0.8$, $\lambda=0.05$ and $V_0\approx
10^{-13}$, and the blue-dotted curve presents the phantom-divide
crossing and corresponds to $\xi=-0.25$, $\lambda=40$ and
$V_0\approx 10^{-12}$. $\lambda$ and $V_0$ are measured in
$\kappa^2$-units and the -1-line is depicted for convenience. }}
\label{plot}
\end{figure}

In Fig.~\ref{plot} we depict the $w_{DE}$-evolution for three
realizations of the scenario at hand. In the case of the
black-solid curve the teleparallel dark energy behaves like
quintessence, in the red-dashed curve it behaves like a phantom,
while in blue-dotted curve the dark energy exhibits the
phantom-divide crossing during the evolution.
Note that the crossing behavior in Fig.~\ref{plot} is the one favored by
the observational data, in contrast with viable $f(R)$-gravity
models where it is the opposite one \cite{BGL}.
We remark that in
the above graphs we focus on their qualitative features,
and in particular we maintain the same potential just to stress
that in principle one can obtain the various behaviors with the
same potential. Clearly, one could be quantitatively more accurate
and impose the observational $w_{DE}(z)$ as an input,
reconstructing the corresponding potential. However in the present
work we desire to remain as general as possible.

\section{Discussion-Conclusions} \label{conclusions}

In the present scenario of ``teleparallel'' dark energy
we have added a scalar
field to the Teleparallel Equivalent to General Relativity (TEGR), allowing
for a nonminimal coupling between the field and gravity. In the
minimally-coupled case the cosmological equations coincide with those of
the standard quintessence. However when the nonminimal coupling is switched on
the resulting theory exhibits different behavior. In particular, although
the scalar field is canonical, one can obtain a dark energy sector being
quintessence-like, phantom-like, or experiencing the phantom-divide
crossing during evolution, a behavior that is much richer comparing to
General Relativity (GR) with a scalar field. Moreover, the fact that the
phantom regime can be described without the need of phantom fields, which
have ambiguous quantum behavior \cite{Cline:2003gs}, is a significant
advantage.

The physical reason for the aforementioned difference, despite the
equivalence of pure GR and pure TEGR, is that while in GR one couples the
scalar field with the only suitable gravitational scalar, namely the Ricci
scalar $R$, in the later one couples the scalar field with the only
suitable gravitational scalar, namely the torsion scalar $T$. The richness
of the resulting theory comparing to GR quintessence is additionally
manifested in the fact that, although in the later one can perform a
conformal transformation and transit to an ``equivalent'',
minimally-coupled, theory with transformed field and potential
\cite{nonminimal}, in the former such a transformation does not exist
since one obtains extra terms depending on the torsion tensor itself, as
can be easily verified transforming the vierbeins as
$e^\mu_A\rightarrow \Omega\tilde{e}^\mu_A$ (one
applies in our case the similar analysis of \cite{Yang:2010ji} of the
case of $f(T)$
scenarios). Thus, teleparallel dark
energy cannot be transformed to an
``equivalent'' minimally coupled form, which is known to be able to
describe only the quintessence regime, and this indicates its richer
structure. Such an absence of conformal transformation exists in other
cosmological scenarios too, for example in scalar-field models with
non-minimal derivative couplings, where it is also known that the resulting
theories possess a richer structure \cite{Amendola:1993uh}.

The addition of a scalar field to TEGR was inspired by the corresponding
procedure in GR. However, although in GR one can alternatively and
equivalently generalize the action to $f(R)$, freeing himself of the need
to add the scalar field, in the teleparallel formulation of GR the
generalization to $f(T)$ \cite{Bengochea:2008gz} seems to spoil the local
Lorentz invariance for all functions apart from the linear one
\cite{Li:2010cg}. However, at the background level no new degrees of
freedom are present, while at linear perturbation the new vector degree of
freedom only satisfies constraint
equations
\cite{Li:2011wu}.
Similarly, in our generalization of TEGR, in the case of non-minimal
coupling, a Lorentz-violating term appears (the last term in the left hand
side of (\ref{eom2})), despite the fact that the theory is linear in $T$.
However, no new degree of freedom will appear at the
background level on which we focus on this work. Clearly, going beyond
background evolution and examine whether the Lorenz violations do indeed
appear under cosmological geometries and scales (we have checked that at
the low-energy limit, the theory's basic
Parametrized Post Newtonian parameters are consistent with Solar System
observations), and if they can be
detected, is an interesting and open subject, as it is in
$f(T)$ gravity too, and will be incorporated in more details elsewhere.

In summary, the rich behavior of teleparallel dark energy makes it a
promising cosmological scenario. In this work we have desired to remain as
general as possible, and present its basic and novel features. Clearly,
before it can be considered as a good candidate for the description of
nature, one needs to investigate various subjects, such as to perform a
detailed perturbation analysis, to use observational data in order to
constrain the parameters of the model, to examine the phase-space
behavior in order to reveal the late-time cosmological features, etc.
Such aspects, although necessary, lie outside the goal of the present work
and are left for future investigations.

{\vspace{0.1cm}}
{\bf{acknowledgments}}\\
The work was supported in part by National Center of Theoretical Science
and  National Science Council
(NSC-98-2112-M-007-008-MY3)
of R.O.C.

\end{document}